\documentclass[runningheads]{llncs}

\usepackage{hyperref}

\usepackage{cleveref}

\usepackage{subcaption}

\usepackage[T1]{fontenc}
\usepackage{graphicx}
\begin{document}
%
\title{Controlling Variability in Neuroimaging Analysis Pipelines with Software Product Lines}

\authorrunning{G. LE BARTZ LYAN et. al}
 \author{Gauthier LE BARTZ LYAN\inst{1}\orcidID{0000-0003-4711-6636} \and 
Paul TEMPLE\inst{1}\orcidID{0000-0002-8276-0593} \and 
Boris CLENET\inst{1}\orcidID{0009-0006-7727-5100} \and 
Mathieu ACHER\inst{1}\orcidID{0000-0003-1483-3858}}
%
 \institute{Univ Rennes, Inria, CNRS, IRISA}

\maketitle              
\begin{abstract}

Magnetic resonance imaging (MRI) is an acquisition technique used for medical diagnosis.
It is notably used for structural and functional MRI (or fMRI) image analysis.
MRI data is known to be complex and noisy, so pipelines are created to extract and enhance relevant information in retrieved images.
The NARPS study highlighted the significant variability that can emerge from different pipelines set up by independent teams while trying to answer the same research question about brain functioning.
Having different implementations realizing the same task is a typical playground for software product lines (SPLs), which may be out of the sight of neuroimage practitioners.
Thus, proposing an SPL out of these different pipelines falls back on a typical reverse-engineering problem with its usual difficulties.
In this document, we advocate for the conceptual adoption of SPL principles as a promising direction to improve reproducibility and consistency in neuroimaging research.

\end{abstract}
\section{Introduction}

Magnetic Resonance Imaging (MRI), a non-invasive medical imaging technique, has highly improved medical diagnosis as a complement to X-ray and other kinds of analyses.
MRI is used mostly for two types of brain studies: structural analysis and functional analysis. Structural analysis involves combining images from one or multiple subjects to highlight specific anatomical regions of the brain. Functional analysis, on the other hand, aggregates brain images over time, introducing a temporal dimension. This allows researchers to identify regions of the brain that become active (or inactive) in response to specific events. For example, functional MRI (fMRI) can reveal what happens in the brain when a human subject is asked to answer a question or move a part of their body.
Despite many advantages, MRI data present significant challenges in terms of analysis and interpretation. A variety of noise sources can compromise data quality, including biological variability (such as differences in brain structure between individuals), subject movement during acquisition, and technical artifacts arising from image processing, such as interpolation or rescaling.

To extract meaningful signals from complex MRI data and minimize noise, neuroscientists rely on software suites that assemble different components that we call pipelines. However, no formal conventions exist to describe these pipelines, despite
initiatives like COBIDAS~\cite{nichols2017best} that are promoting standardized best practices for analysis.
The NARPS study conducted by Botvinik-Nezer et al.~\cite{botviniknezer:inserm-02914443} investigated how different research teams process a single fMRI dataset in response to a set of closed binary research questions (\textit{i.e.}, with possible answers being only 'yes' or 'no'). The pipelines were unconstrained, letting research teams use their usual software suites and create the pipeline they want. The NARPS study revealed that even when using the same data and hypotheses, widely used software tools can produce non-consistent results.
This means that, when aggregating the different results to answer the research questions, observations could lead to half the research teams answering 'yes' while the other half would answer 'no'.
These concerns were later reinforced by Bowring et al.~\cite{Bowring}.

In a similar vein, Germani~\cite{germani:tel-04809662} examined the analytical variability arising from different processing methods and proposed several improvements to enhance the validity of neuroimaging studies. Vila et al.~\cite{10.1145/3641525.3663626} demonstrated that variations in computing hardware can significantly affect the output of neuroimaging pipelines, showing that the same analysis can yield different results when run on different systems. Chatelain et al.~\cite{10.1109/TC.2024.3475586} investigated numerical variability in FMRIPrep~\cite{esteban2019fmriprep,esteban2020analysis}, a widely used data preprocessing framework, and showed that changes in its configuration can alter the final preprocessed images. Finally, Sokolowski et al.~\cite{sokolowski2024impact} explored how software versioning impacts brain structural analysis. They revealed that different versions of the same software can produce statistically significant discrepancies in the results.

\begin{figure}[htbp]
   \centering
   \includegraphics[width=\textwidth]{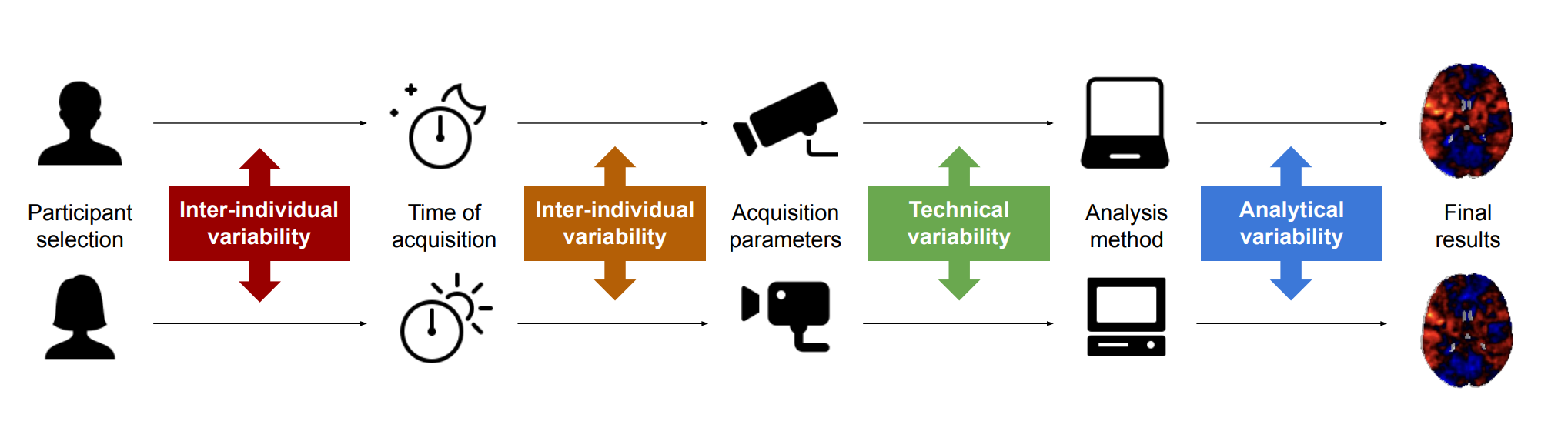}
   \caption{Different sources of variability in neuroimaging studies as shown by Germani~\cite{germani:tel-04809662}, Licensed under CC BY 4.0}
   \label{fig:var}
\end{figure}

With so many layers of variability as shown in Figure~\ref{fig:var}, it is essential to understand what happens in each of them. To achieve this, establishing a mapping of the impact of each layer is a necessary first step. This requires the ability to explore every layer using appropriate protocols and tools. 
In particular, the layer representing the ability to choose between different software is one of the most challenging. These software have no codebase in common or do not necessarily share coding conventions, thus reasoning and associating the realization of different functionalities is difficult.
In addition, once the association is done, comparing the same functionality in different software requires numerous dedicated pipelines, extensive runs, and the generation of large volumes of data.

In this paper, we demonstrate that NARPS Open Pipelines, an open source pipeline suite, exhibits SPL-like characteristics and supports a structured variability analysis. While we do not formally define the SPL architecture, our observations suggest shared components, variation points, and modular structure, key hallmarks of SPLs.
We share our experience in trying to replicate pipelines from NARPS and exhibit the challenges we faced that could be alleviated by moving to a SPL structure. We also propose discussions about the specificities of SPL in the current neuro-imaging world.
Furthermore, neuro-imaging could be a playground to further push the exploration of layers of variability as proposed by Lesoil~\cite{lesoil2021deep}.

The rest of the paper is structured as follows: first, we explain how neuroimaging pipelines are designed and highlight the room for variability management in Section~\ref{sec:background}. Then, Section~\ref{sec:narps} reports our experience in replicating some of the pipelines from the NARPS study and the challenges we have faced. Section~\ref{sec:conclusion} summarizes our experience and draws some future directions and challenges to adopt a variability-aware strategy in the development of neuroimaging pipelines.

\section{Background}
\label{sec:background}

\subsection{Neuroimaging software} 

Nipype~\cite{nipype} is a widely used middleware for neuroimaging workflow management. It aims at reducing software heterogeneity and encapsulating domain-specific processing concepts. By providing a standardized interface for integrating diverse tools (e.g., FSL, SPM, AFNI), Nipype enables researchers to focus on data flow and analysis design rather than underlying software architecture or version-specific dependencies. However, seamlessly switching between different neuroimaging tools or workflows remains challenging without requiring significant manual intervention, including rewriting all or part of the pipelines and modifying the runtime environment~\cite{maumet2025variability,youenn}. 
This harms the code quality of workflows and limits the ability to explore the pipelines' configuration space.
In the meantime, these pipelines are designed to process similar data and provide answers to research questions defined by neuroimaging practitioners. Hence, our interest in using software product lines to explore this context.

\subsection{Software Product Lines}
With the mass customization industry, the software engineering development process has changed from developing a single piece of software for specific usages of a particular customer to developing multiple similar software systems that share a common code base and address multiple customers' usages at once, the so-called Software Product Lines (SPL).
This way, reusing becomes a central concern of software development.
Pohl \textit{et al.} \cite{pohl2005software} argue that reuse as much as possible brings (1) a reduced development cost, since less code is written; mostly glue code has to be created which is usually smaller than developing a complete functionality; (2) a reduced time-to-market since the product is not developed from scratch; (3) an improved code quality, since with the reusability concern in mind, newly developed code should be modularized. Thus easing testing, readability and reducing the complexity and technical debt of future code.

Previous industrial success stories have been reported in the SPLC conference\footnote{\url{https://splc.net/fame.html}}. These companies developed multiple products for different needs but wanted to capitalize on previous work. They were aware that complete parts of their products were generic and only some other parts would need custom code for customers' specific needs. Thus, based on the multiple already developed products, they tried to reverse engineer what would be the structure of their software product lines.
We think that neuroimaging is in a similar context, especially with tools like Nypipe that simplify the development of pipelines by interfacing different tools. However, the different pipelines, which make up the SPL, would not be owned by a unique entity as it is the case in industrial success stories mentioned before.

\subsection{Potential impacts for neuroimaging research and their pipelines}
The growing body of evidence on variability in neuroimaging~\cite{botviniknezer:inserm-02914443,Bowring} shows that differences in software, versions, and environments can lead to divergent results.
Yet, as the pipelines and the tools are built differently, performing comparisons seems difficult. Ablation studies seem impossible as each tool is intricated with another, the definitions of functions are specific to each tool, etc.
However, some strategies remain, and from a high-level view, some functionalities or successions of functions can be identified (e.g., noise reduction or blur), which suggests that software product lines could be interesting to further analyze this variety of pipelines.


Hereafter, we report on our experience replicating neuroimaging pipelines and highlight some challenges that we foresaw.
We would like to remind that this work is exploratory. It is based on a single dataset and a limited set of pipelines SPM12\footnote{\href{https://www.fil.ion.ucl.ac.uk/spm/software/spm12/}{https://github.com/ReproNim/neurodocker}} associated with MATLAB. It does not claim to generalize across all neuroimaging tools or software stacks. Our findings are preliminary and intended to stimulate further research into the application of SPL principles in neuroimaging pipeline design.

\section{NARPS Open Pipelines and pipelines replication}\label{sec:narps}

The NARPS Open pipelines project\footnote{\href{https://github.com/Inria-Empenn/narps\_open\_pipelines}{https://github.com/Inria-Empenn/narps\_open\_pipelines}}, led by the Empenn research team at the University of Rennes, France, is an effort at reproducing the analysis pipelines originally described in the NARPS study within a standardized, reproducible environment, based on Nypipe pipelines and Neurodocker runtimes\footnote{\href{https://github.com/ReproNim/neurodocker}{https://github.com/ReproNim/neurodocker}}. The goal is to enable researchers to execute these pipelines and replicate the results reported in the original study, which are publicly available on NeuroVault~\cite{neurovault}. The project is hosted on GitHub and is fully open-source, making it accessible to the broader neuroimaging community. 

As of late 2025, only a limited number of pipelines have been successfully reproduced. This progress is hindered by the significant challenges of retro-engineering pipelines based on sparse, often informal documentation, primarily recorded in spreadsheet files with limited technical details. The lack of clarity in descriptions, combined with differences in software versions and implementation choices, has led to substantial reinterpretation, development overhead, and testing effort. These challenges emphasize the difficulties of translating ad hoc, team-specific workflows into reproducible, maintainable, and transparent software systems.

Using an ad hoc Dockerfile generator\footnote{\href{https://gitlab.inria.fr/glebartz/narps/-/blob/main/Docker\_gen.ipynb}{https://gitlab.inria.fr/glebartz/narps/-/blob/main/Docker\_gen.ipynb}} inspired from Neurodocker's core Dockerfile\footnote{\href{https://github.com/ReproNim/neurodocker}{https://github.com/ReproNim/neurodocker}}, we successfully ran three pipelines (named after their anonymized teams) from NARPS Open Pipelines\footnote{\href{https://github.com/Inria-Empenn/narps_open_pipelines/tree/main/narps_open/pipelines}{https://github.com/Inria-Empenn/narps\_open\_pipelines/tree/main/narps\_open/pipelines}}, 2T6S, U26C and V55J, all based on the same software stack that uses SPM12 and MATLAB.

We specifically investigated aspects of deep variability in neuroimaging analysis, drawing on recent work in reproducibility \cite{lesoil2021deep,acher2022reproducible,acher2024embracing}. Focusing on preprocessing, image manipulation, and statistical map generation, we conducted controlled experiments by performing mutually exclusive swaps of key software components, including MATLAB, SPM12 release version, and underlying hardware platforms (Intel and AMD x86 CPUs, and Apple ARM architectures). To ensure reproducibility and isolate environmental influences, all tests were executed within dedicated, isolated Docker containers.\footnote{All our results are available in the repository \href{https://gitlab.inria.fr/glebartz/narps}{https://gitlab.inria.fr/glebartz/narps}}. 
Roughly, our replications take raw images as inputs, apply the signal processing steps described in the pipelines, and superimpose the resulting visualization images to detect differences.

\begin{figure}[h]
  \centering
  \includegraphics[width=0.8\textwidth]{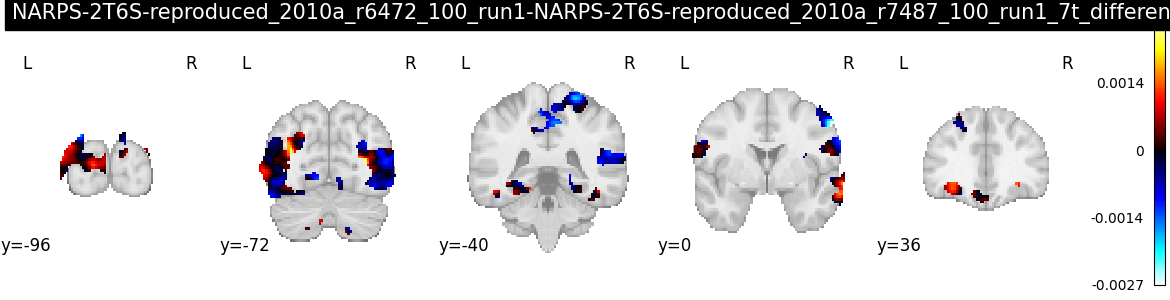}
  \caption{ Example of observed differences between brain activation patterns (NARPS Hypothesis 7, thresholded) from the 2T6S team}
  \label{fig:exp}
\end{figure}

The retrieved results exhibit minor numerical discrepancies in voxel values, ranging from $6e^{-5}$ to $1e^{-2}$. Figure~\ref{fig:exp} is a representative example of the magnitude and importance of discrepancies that were observed. 
These differences are small and seem negligible; they are below typical thresholds used in fMRI statistical analysis (e.g., p < 0.05). 
This suggests a degree of robustness in the application of core operators across different MATLAB versions, and further indicates that variations in hardware platforms, such as transitions between x86 and ARM architectures, do not introduce substantial deviations in computed results.
However, their impact on statistical inference should be evaluated by domain experts, particularly when studying sensitive brain regions, as we are non-experts in this domain.

However, despite their apparent insignificance in the context of final conclusions, such disparities should be considered carefully. The absence of explicit thresholds or domain-specific criteria for determining significance means that these differences should be evaluated by domain experts, who can assess whether such magnitudes are clinically or statistically meaningful. More importantly, highlighting these variations raises awareness among researchers about the sensitivity of (f)MRI processing to environmental factors such as software version, platform architecture, and computational environment. This emphasizes the critical need for standardized, well-documented, and reproducible processing pipelines that explicitly control and report such environmental variables, ensuring both transparency and scientific rigor in neuroimaging analysis. Finally, our empirical analysis of pipeline variability reveals three key SPL characteristics: (1) shared core modules across pipelines; (2) variation points where environmental factors can be configured; and (3) modular structure enabled by containerization.

\subsection{Towards Software Product Lines for (f)MRI analysis}

\begin{figure}[htbp]
   \centering
   \begin{subfigure}{0.48\textwidth}
       \centering
       \includegraphics[scale=0.19]{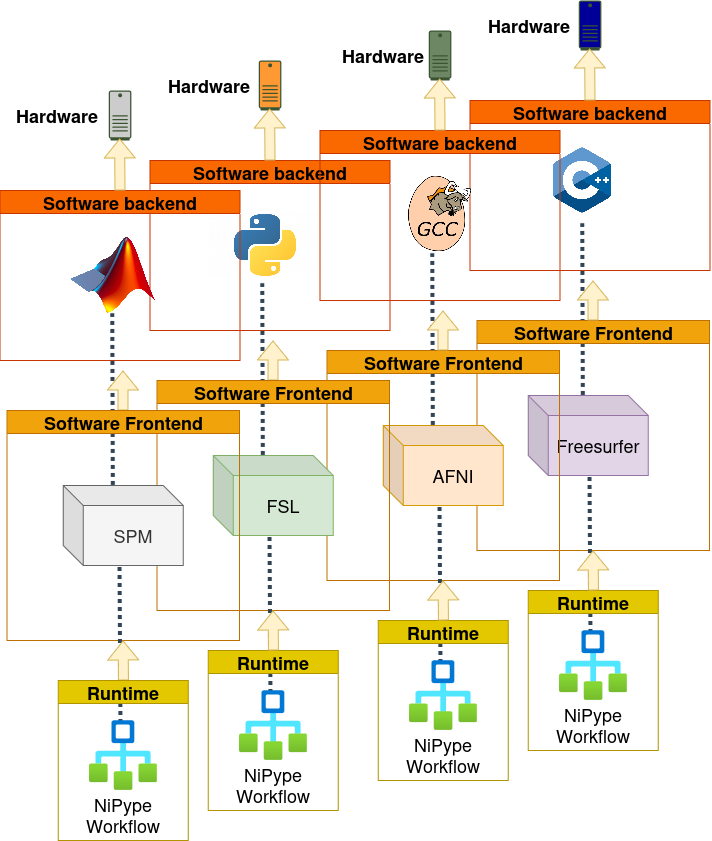}
       \caption{Sample of pipelines as currently done, which limit reusability. Nipype (at the bottom) links to a specific frontend, in turn accessing specific implementations of a backend.} 
       \label{fig:pipeline_count}
   \end{subfigure}
   \hfill
   \begin{subfigure}{0.48\textwidth}
       \centering
       \includegraphics[scale=0.19]{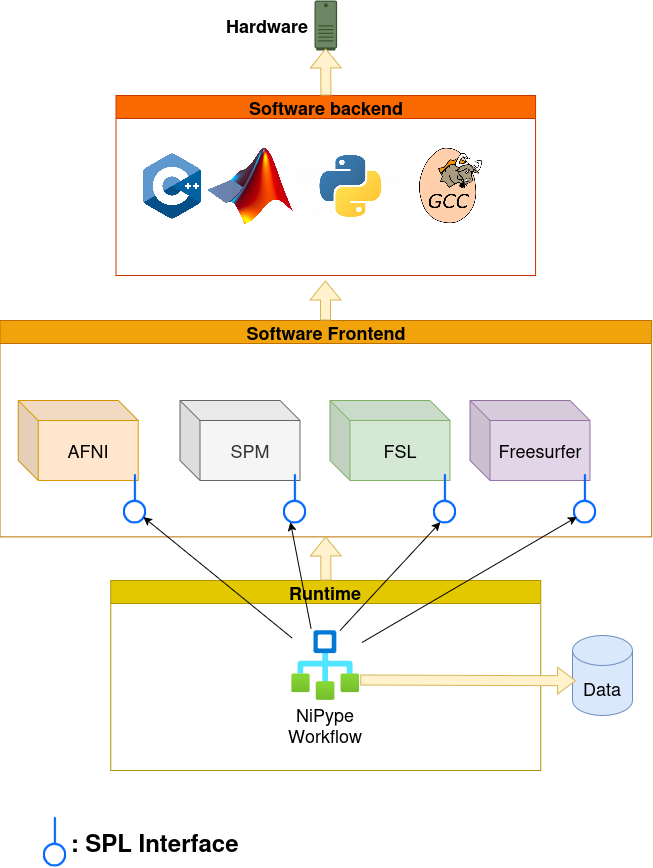}
       \caption{Sample of pipelines with an SPL design in mind. The different frontends can be interchanged easily within a Nipype workflow, and the specific code is pushed as late as possible in the call to specific functions from the software backends.}
       \label{fig:spl}
   \end{subfigure}
   \caption{Current state of neuroimaging with a specific pipeline per tool and hardware (left) VS our conceptual SPL view of neuroimaging pipelines (right).}
\end{figure}

With our current experiment, we have examined only a limited subset of the neuroimaging software ecosystem. SPM (and its MATLAB-based environment) are by no means the only adopted tools within the community. Therefore, generalizing our findings requires a comprehensive evaluation across a broader range of existing neuroimaging software packages. However, as highlighted in Section~\ref{sec:narps}, reproducing and validating processing pipelines across diverse tools is highly resource-intensive. Furthermore, while middleware such as Nipype offers partial integration between tools, it often lacks complete or standardized interfaces for each platform. As a result, dedicated pipelines would need to be developed for every major software suite and testing purposes. Such a comprehensive effort remains infeasible with current resources. Figure~\ref{fig:pipeline_count} illustrates the vast number of distinct pipelines that would be required to test only a representative sample of existing neuroimaging tools, demonstrating the practical limitations of a broad-scale validation approach at present.

One conceptual solution to the fragmentation of neuroimaging pipelines is to introduce a standardized software interface that every neuroimaging tool, whether developed by commercial vendors, open-source communities, or individual researchers, should implement. This interface would enable middleware components to abstract away differences in underlying implementations, facilitating interoperability through a factory pattern. In this model, processing pipelines would make standardized, uniform calls to a single interface, and the specific backend (e.g., FSL, SPM, AFNI, or custom tools) could be swapped simply by changing a single import statement in a Python script. As illustrated in Figure~\ref{fig:spl}, this modular architecture enables dynamic configuration and reuse across different tools.

Such an approach would bring to neuroimaging the “Code Once, Run Anywhere” (CORA) paradigm, a concept famously realized in Java through the Java Virtual Machine, allowing researchers to develop and execute pipelines with consistent behaviors, regardless of the underlying software suite. This not only enhances reproducibility and portability but also provides a scalable foundation for managing variability, promoting interoperability, and enabling transparent, plug-and-play analysis workflows.


\subsection{Challenges}

The main problem to adopt such an approach is that all stakeholders providing software within the neuroimaging community need to reconsider how their tools are designed and made accessible to external users. This implies the need for a standardized, well-defined interface that specifies the core capabilities a neuroimaging software component must offer to support consistent and interoperable MRI data processing. Such an interface, which would outline mandatory features, data formats, and functional requirements, could be established through a collaborative effort involving software developers, researchers, and domain experts. It might emerge as a community-led initiative, a consortium of providers, or, initially, as a proposal from a single entity offering a pilot interface. Crucially, given the diversity and autonomy of current stakeholders, a compelling first step would be a proof-of-concept demonstrating the feasibility and benefits of such a standardized interface. A successful, publicly shared proof-of-concept would serve as a catalyst to build community engagement, foster trust, and establish a foundation for broader adoption.

\section{Conclusion}
\label{sec:conclusion}

The analysis of variability in neuroimaging pipelines, particularly as revealed by the NARPS study, demonstrates that current practices, while functionally effective, lack structural coherence and reproducibility. Despite the diversity of tools and workflows, the shared architectural patterns and configurable components across pipelines suggest a natural alignment with the principles of Software Product Line (SPL) engineering.

Our empirical exploration of the NARPS open pipelines project, including controlled experiments on software versions, hardware platforms, and processing configurations, reveals that while minor numerical differences occur, these are often within clinically and statistically insignificant bounds and are most probably absorbed by downstream analysis. This highlights both the robustness of core processing operations and the critical need for greater transparency and control over environmental variables.

However, the absence of standardized interfaces and interoperable middleware remains a major barrier to consistent, reproducible, and scalable pipeline development. To address this, we advocate for the adoption of an SPL-inspired framework, one that enables a “Code Once, Run Anywhere” (CORA) paradigm through a shared, well-defined interface for neuroimaging software components. This would allow researchers to build reusable, configurable pipelines that dynamically plug in different backends (e.g., FSL, SPM, AFNI, FreeSurfer) without requiring extensive rework or manual intervention.

Implementing such a vision requires a collaborative, community-driven initiative to establish a standardized software interface, supported by a proof-of-concept demonstrating feasibility and benefits. While technical and organizational challenges remain, the potential gains in reproducibility, transparency, and scientific rigor make this a compelling and necessary step forward.


%
%
 \bibliographystyle{splncs04}
 \bibliography{biblio}
 
\end{document}